\documentclass[twocolumn]{revtex4}
\usepackage[dvips]{graphicx}
\begin{document}

\title{
Impurity-induced transition and impurity-enhanced thermopower 
in the thermoelectric oxide NaCo$_{2-x}$Cu$_x$O$_4$
}

\author{I. Terasaki}
\email{terra@mn.waseda.ac.jp}
\affiliation{
Department of Applied Physics, Waseda University, Tokyo 169-8555, Japan\\
Precursory Research for Embryonic Science and
Technology, Japan Science Technology, Tokyo 108-0075, Japan
}

\author{I. Tsukada}
\affiliation{
Central Research Institute for Electric Power Industry,
Komae 201-8511, Japan
}

\author{Y. Iguchi}
\affiliation{
Department of Applied Physics, Waseda University, Tokyo 169-8555, Japan
}

\date{\today}

\begin{abstract}
 Various physical quantities are measured and analysed for
 the Cu-substituted thermoelectric oxide NaCo$_{2-x}$Cu$_x$O$_4$.
 As was previously known, the substituted Cu enhances the 
 thermoelectric power, while it does not increase the resistivity 
 significantly.
 The susceptibility and the electron specific-heat are substantially
 decreased with increasing $x$, which implies that the substituted Cu
 decreases the effective-mass enhancement.
 Through a quantitative comparison with the heavy fermion compounds
 and the valence fluctuation systems, we have found that
 the Cu substitution effectively increases the coupling between
 the conduction electron and the magnetic fluctuation.
 The Cu substitution induces a phase transition at 22 K
 that is very similar to a spin-density-wave transition. 
\end{abstract}

\pacs{}

\maketitle

\section{Introduction}
Recently layered cobalt oxides have been extensively investigated as a
promising candidate for a thermoelectric material. 
The thermoelectric material is a material that shows large thermopower ($S$),
low resistivity ($\rho$) and low thermal conductivity ($\kappa$)
\cite{mahan}, and  
a quantity of $Z \equiv S^2/\rho\kappa$ called the figure of merit
characterizes the thermoelectric conversion efficiency.
Thermoelectric device can generate electric power from heat through the 
Seebeck effect, and can pump heat through the Peltier effect.
Thus far oxides have been regarded as unsuitable for thermoelectric
application because of their poor mobility, but some years ago
Terasaki {\it et al.} found that a single crystal of the layered oxide
NaCo$_2$O$_4$ exhibits good thermoelectric performance \cite{terra}.
Fujita {\it et al.} showed that 
the dimensionless figure of merit $ZT$ of a NaCo$_2$O$_4$ single crystal
exceeds unity at $T=$1000 K \cite{fujita},
Ohtaki {\it et al.} \cite{ohtaki} measured $ZT \sim 0.8$ at 1000 K
even in the polycrystalline samples of NaCo$_2$O$_4$.
Thus this compound is quite promising for
thermoelectric power generation at high temperature.

Following NaCo$_2$O$_4$, other layered cobalt oxides, 
Ca-Co-O \cite{li,miyazaki,masset,funahashi}, 
Bi-Sr-Co-O \cite{itoh,itoh2,funahashi2}, 
Tl-Sr-Co-O \cite{hebert} 
have been found to show good thermoelectric performance. 
In particular, Funahashi {\it et al}. \cite{funahashi} 
showed $ZT > 1$  at 1000 K for Ca-Co-O. 
The most important feature is that 
the CdI$_2$ type triangular CoO$_2$ block 
is common to the layered cobalt oxides.
We have proposed that the high thermoelectric performance 
of the layered cobalt oxides cannot be explained 
by a conventional band picture based on the one-electron approximation,
but is understood in terms of the strong electron-electron 
correlation effects,
similarly to the case of heavy-fermion compounds.
In fact the material dependence of the thermopower
quite resembles that of Ce$M_2X_2$  \cite{terra2}.

A prime example for the difficulties of the one-electron picture
is observed in the Cu substitution effects in NaCo$_2$O$_4$ \cite{terra3}.
The thermopower of NaCo$_{2-x}$Cu$_x$O$_4$ is significantly enhanced, 
while the resistivity is nearly independent of $x$.
This is quite surprising in comparison with
normal impurity effects in a metal.
The doped impurity acts as a scattering center in usual cases,
and does not make a significant change in thermopower,
because it is a quantity of the zero-th order of scattering time.
Indeed this is what was observed in high-temperature 
superconductors \cite{tallon}.
Importantly, correlation effects can explain
the large impurity effect on the thermopower,
similarly to the case of dilute magnetic alloys \cite{fischer}.
In this paper, we report on measurement of specific heat, 
susceptibility, Hall coefficient, and transverse magnetoresistance
for NaCo$_{2-x}$Cu$_x$O$_4$ polycrystalline samples,
and discuss the Cu substitution effects quantitatively.

\section{Experimental}
Polycrystalline samples of 
Na$_{1.2}$Co$_{2-x}$Cu$_x$O$_4$ 
($x$=0, 0.1, 0.2 and 0.3) were prepared 
through a solid-state reaction. 
A stoichiometric amount of Na$_2$CO$_3$, Co$_3$O$_4$ and CuO
was mixed and calcined at 860$^{\circ}$C for 12 h in air. 
The product was finely ground, pressed into a pellet, 
and sintered at 920$^{\circ}$C for 12 h in air. 
Since Na tends to evaporate during calcination, 
we added 20 \% excess Na. 
Namely we expected samples of the nominal 
composition of Na$_{1.2}$Co$_{2-x}$Cu$_x$O$_4$ to be 
NaCo$_{2-x}$Cu$_x$O$_4$.

The x-ray diffraction (XRD) was measured 
using a standard diffractometer with Fe K$_{\alpha}$ radiation 
as an x-ray source in the $\theta -2\theta$ scan mode.
The resistivity was measured through a four-terminal method,
and the thermopower was measured using a steady-state technique 
with a typical temperature gradient of 0.5 K/cm.
The Hall coefficient ($R_H$) and the transverse magnetoresistance 
were measured from 15 to 100 K in a closed refrigerator 
inserted into a room temperature bore of a liquid-He free 
superconducting magnet. 
To eliminate the unwanted voltage arising from 
the misalignment of the voltage pads, 
the magnetic field was swept from -7 to 7 T 
with a typical period of 20 min at constant temperatures 
with a stability of 10 mK.
The specific heat was measured using a standard relaxation
method with a mechanical heat switch. 
The mass of the samples used for the measurement 
is typically 1000 mg and the heat capacity 
of the samples is always more than two
orders of magnitude larger than the addenda heat capacity.
The susceptibility was measured with a SQUID susceptometer
in a magnetic field of 1 T.

\begin{figure}
 \begin{center}
  \includegraphics[width=8cm,clip]{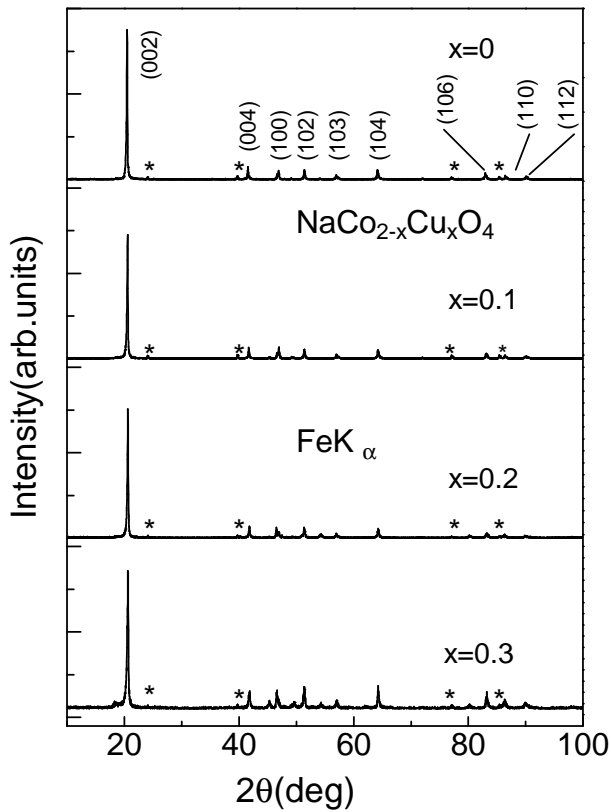}
 \end{center}
 \caption{The x-ray diffraction patterns of the 
 polycrystalline samples of NaCo$_{2-x}$Cu$_x$O$_4$.
 The Fe K$_{\alpha}$ is used a an x-ray source.}
 \label{fig1}
\end{figure}

\section{Results}
Figure 1(a) shows the x-ray diffraction patterns of the prepared samples
of NaCo$_{2-x}$Cu$_x$O$_4$.
Almost all the peaks are indexed as the $\gamma$ phase
\cite{JH,fouassier}, though
a small amount (approximately less than 5\%) of 
unreacted Co$_3$O$_4$ is observed. 
With increasing Cu content $x$, no additional peak appears,
with the patterns unchanged,
which shows that Cu is substituted for Co.
However, the sample of $x$=0.3 shows a higher background noise,
indicating that the crystal quality becomes worse,
possibly owing to the limit of solid solution with Cu.

Figure 2(a) shows the temperature dependence of the resistivity 
for NaCo$_{2-x}$Cu$_x$O$_4$.
All the samples are metallic down to 4.2 K without any indication
of localization.
It should be noted that the increased resistance due to 
the substituted Cu is of the order of 10$\mu\Omega$cm for 1 at.\% Cu,
which is anomalously small in the layered 
transition-metal oxides \cite{fukuzumi}.
Another important feature is that the resistivity for the Cu substituted
samples show a kink near 22 K as indicated by the dotted line.
Since the temperature dependence is steeper below 22 K, 
the density of states (or the carrier concentration) decreases below 22 K.

\begin{figure}[b]
 \begin{center}
  \includegraphics[width=8cm,clip]{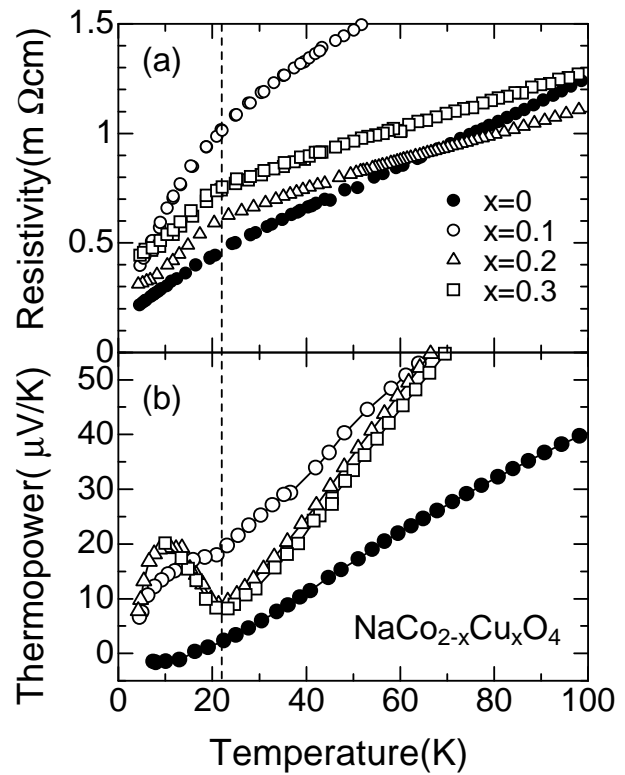}
 \end{center}
 \caption{(a) The resistivity and (b) the thermopower
 of polycrystalline samples of NaCo$_{2-x}$Cu$_x$O$_4$.}
 \label{fig2}
\end{figure}

Figure 2(b) shows the temperature dependence of the thermopower
for NaCo$_{2-x}$Cu$_x$O$_4$.
Thermopower increases with increasing $x$ with 
a dip near 22 K and a peak near 10-15 K.
As we previously shown in the analysis for Bi-Sr-Co-O \cite{itoh2},
a low-temperature thermopower ($S$) of the layered Co oxides 
is determined by the diffusive term that 
is proportional to temperature ($T$).
Accordingly $S/T$ is an essential parameter
similarly to the electron specific-heat coefficient.
Thus, the dip, rather than the peak, is a meaningful temperature,
which corresponds to the onset of the enhancement 
in $S/T$ at low temperatures.
It should be emphasized that the dip temperature is 
nearly the same as the kink temperature for the resistivity,
which strongly suggests that this temperature is 
related to a kind of phase transition.

\begin{figure}
 \begin{center}
  \includegraphics[width=8cm,clip]{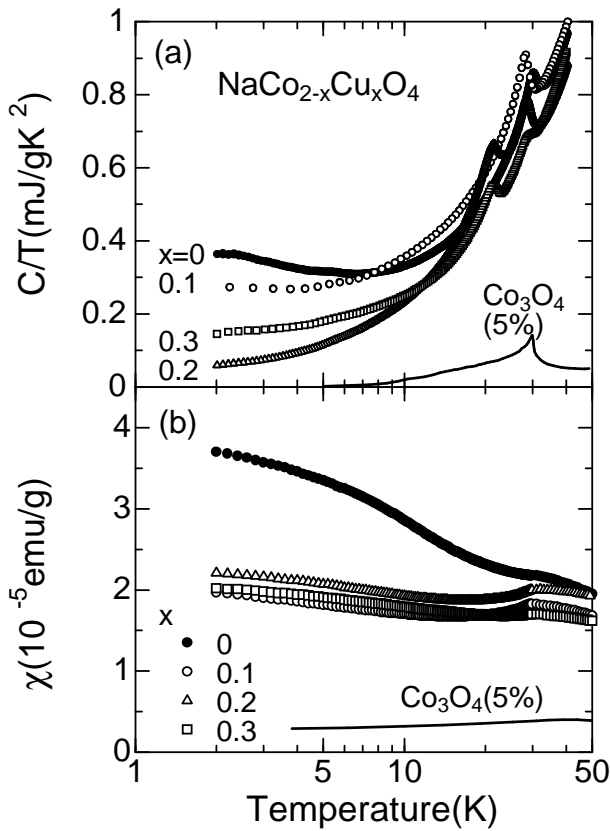}
 \end{center}
 \caption{(a) The specific heat $C$ and 
 (b) the susceptibility $\chi$
 of polycrystalline samples of NaCo$_{2-x}$Cu$_x$O$_4$.
 Note that $C/T$ is plotted in order to emphasize the 
 electron specific heat coefficient $\gamma$.
 The samples include 5\% of unreacted Co$_3$O$_4$,
 and the data for 5\% Co$_3$O$_4$ taken from 
 Refs, \cite{LB1,LB2} 
 are plotted by the solid curves.}
 \label{fig3}
\end{figure}

Figure 3(a) shows the specific heat $C$ for NaCo$_{2-x}$Cu$_x$O$_4$.
In order to emphasize the $T$-linear electron specific heat,
we plot $C/T$ as a function of temperature.
As shown in Fig. 3(a), the $C/T$ value at 2 K decreases 
with increasing the Cu content from 0 to 0.2,
which means the decrease of the electron specific heat 
coefficient $\gamma$ with $x$.
(For the sample of $x$=0.3, the $C/T$ value increases again, 
which might be due to an extrinsic origin such as
the solid-solution limit of Cu.)
Since $\gamma$ is proportional to the density of states 
and the mass-enhancement factor,
the present results indicate that the either or both decrease 
with the Cu substitution.
As for high temperatures, all the data show a peak near 30 K
which is the antiferromagnetic transition of the unreacted 
Co$_3$O$_4$, as shown by the solid curve \cite{LB1}.
As mentioned above, the x-ray diffraction patterns reveal
less than 5 at.\% of unreacted Co$_3$O$_4$,
which is consistent with the peak height of the specific heat at 30 K. 
We should  emphasize here that the existence of Co$_3$O$_4$
does not seriously affect the estimation of $\gamma$,
because the $C/T$ value for Co$_3$O$_4$ is negligibly small 
at low temperatures.
For $x=$0.2 and 0.3, another peak appears in 
the specific heat near 22 K,
which is close to the kink temperature in $\rho$,
and the dip temperature in $S$.
We thus conclude that the 22-K anomaly comes from
a (2nd order) phase transition.

Figure 3(b) shows the susceptibility ($\chi$) of NaCo$_{2-x}$Cu$_x$O$_4$.
The substituted Cu also decreases the susceptibility,
indicating the decrease of the density of states and/or 
the mass-enhancement factor.
A broad hump near 30 K is due to the antiferromagnetic 
transition of the unreacted Co$_3$O$_4$,
as shown by the solid curve \cite{LB2}.
Interestingly, there is no anomaly near 22 K in the susceptibility,
suggesting that the transition at 22 K 
is not the magnetic transition of impurity phases.
We further note that the Curie-like contribution
is absent in the susceptibility at low temperatures,
which shows that magnetic impurities 
are unlikely to exist other than Co$_3$O$_4$.
Quantitatively, the decrease of $\chi$ by Cu 
is more moderate than that of $C/T$.
$C/T$ decreases by a factor of ten from $x=0$ to 0.2, 
whereas $\chi$ decreases only by a factor of two.
This implies that the 22-K transition causes a dramatic reduction
of the electron entropy possibly owing to a (pseudo)gap formation, 
while it does not alter the magnetic  excitation at $k=0$.
The nature of the 22-K transition will be discussed in the next section.

\begin{figure}
 \begin{center}
  \includegraphics[width=8cm,clip]{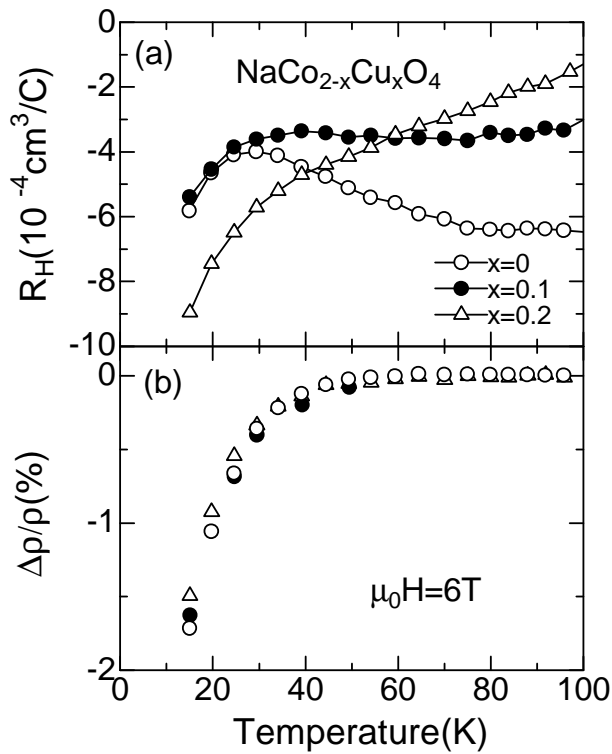}
 \end{center}
 \caption{(a) The Hall coefficient $R_H$ and
 (b) the magnetoresistance $\Delta\rho/\rho$
 of polycrystalline samples of NaCo$_{2-x}$Cu$_x$O$_4$.}
 \label{fig4}
\end{figure}

Figure 4(a) shows the Hall coefficient ($R_H$) of NaCo$_{2-x}$Cu$_x$O$_4$.
The sign is negative below 100 K, and the magnitude is 
as small as 4-6$\times$10$^{-4}$ cm$^3$/C.
The Cu substitution does not change the magnitude so much, 
indicating that the carrier concentration is 
nearly unchanged. 
By contrast, it changes the temperature dependence 
in a complicated way, which implies
that plural kinds of carriers are 
responsible for the electric conduction.
The band calculation by Singh \cite{singh} 
reveals that the two bands of different symmetries cross 
the Fermi level for NaCo$_2$O$_4$.
(See the next section)

Contrary to the Hall effect, 
the magnetoresistance is weakly dependent on the Cu substitution,
as shown in Fig. 4(b).
By taking a closer look at the $x$ dependence,
one can see that the negative magnetoresistance is 
gradually suppressed by the substituted Cu.
This implies that the kink of the resistivity is more or less  
smeared against magnetic field,  which suggests that
the magnetic field suppresses the 22-K transition.

\section{Discussion}
Before going into details,
we will begin with a brief review  on
the electronic states of NaCo$_2$O$_4$.
As is well known, the five-fold degenerate $d$ orbitals
split into two-fold ($e_g$) and three-fold
($t_{2g}$) degenerate levels in an oxygen octahedron. 
In the real triangular CoO$_2$ block, the octahedron is compressed
along the $c$ direction, and the degeneracy is further quenched,
where the lower $t_{2g}$ levels split into $e_g$ and $a_{1g}$ levels. 
The lower $e_g$ levels correspond to the orbital spread along the
CoO$_2$ block to make a relatively broad band,
whilst the $a_{1g}$ orbital is spread along the $c$ direction
to make a nearly localized narrow band.
Since each Co ion is 3.5+ with $(3d)^{5.5}$,
the highest occupied orbital is  $a_{1g}$ in the atomic limit,
and the main part of the Fermi surface consists of 
the narrow $a_{1g}$ band.
In the real band calculation, however, 
there is significant hybridization 
between the $a_{1g}$ and $e_g$ levels, and
the broader $e_g+a_{1g}$ band touches the Fermi level 
to make small Fermi surfaces \cite{singh}.

We have proposed that the electronic structure of NaCo$_2$O$_4$
is similar to that of the Ce-based intermetallics,
a prime example of valence-fluctuation/heavy-fermion 
compounds \cite{terra2}.
$\gamma$ and $\chi$ of NaCo$_2$O$_4$ are as large as those of CePd$_3$.
In this context, the large thermopower of NaCo$_2$O$_4$
is explained in terms of a diffusive contribution of a metal 
with a heavily enhanced effective mass,
and is roughly proportional to $\gamma$.
In the heavy fermion compounds, the broad conduction band 
crosses the Fermi energy, and the narrow (localized) f band 
is located below the Fermi energy. 
For NaCo$_2$O$_4$, the broad $e_g+a_{1g}$ band and 
the narrow $a_{1g}$ band do exist, 
but both of which cross the Fermi energy 
to form two kinds of the Fermi surfaces. 
Thus it is not trivial whether or not the two Fermi surfaces
behave heavy-fermion-like in the charge transport.
At least we can say that the two Fermi surfaces play different 
roles, where the Cu substitution induces different effects:
$\rho$ is weakly dependent on the Cu content $x$,
whereas $S$, $\chi$ and $\gamma$ are strongly dependent on $x$.
$S$, $\chi$ and $\gamma$
are basically proportional to the density of states in the lowest order,
which are determined by the large Fermi surface of the $a_{1g}$ symmetry.
In contrast, the carriers on the $e_g+a_{1g}$ band can be highly mobile,
because it is spread along the in-plane direction. 
In short, the $a_{1g}$ and $e_g+a_{1g}$ bands 
are responsible for the large thermopower 
and a good electric conduction, respectively.

Existence of the $a_{1g}$ and $e_g+a_{1g}$ bands was 
suggested from the angular dependence of the X-ray absorption 
spectroscopy experiment \cite{mizokawa, mizokawa2},
where the valence bands of NaCo$_2$O$_4$ consist of
the $a_{1g}$ and $e_g+a_{1g}$ bands.
The valence band of the less conductive Bi-Sr-Co-O 
is mainly composed only of the $a_{1g}$ band,
which is consistent with our speculation that the $e_g+a_{1g}$ 
band is responsible for the metallic nature of NaCo$_2$O$_4$.
The large Fermi surface suggested by the band calculation
was not seen in the angle photoemission
spectra for Bi-Sr-Co-O, which indicates that 
the band calculation should be modified by additional
effects such as the electron-electron or electron-phonon effects.

\begin{figure}
 \begin{center}
  \includegraphics[width=8cm,clip]{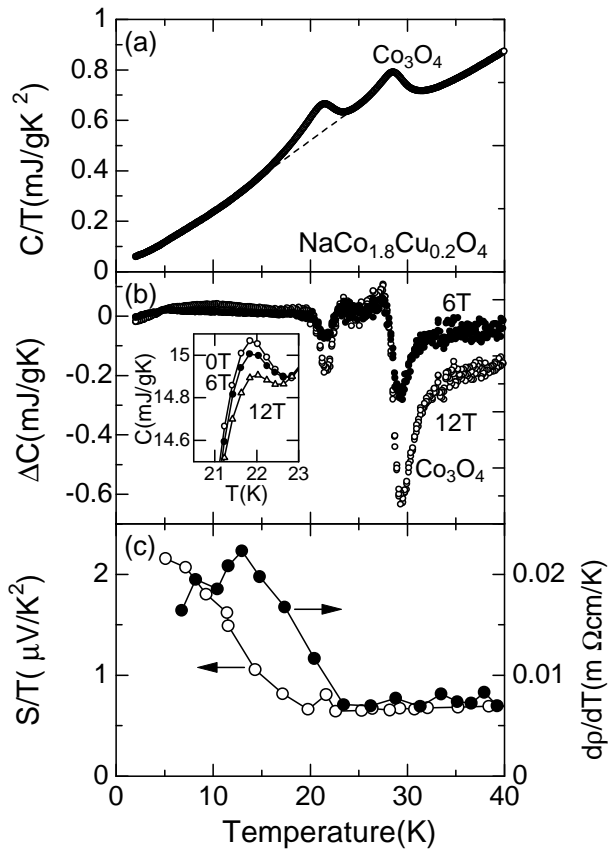}
 \end{center}
 \caption{Phase transition at 22 K for the Cu-substituted sample
 ($x$=0.2). (a) Specific heat, (b) magneto-specific heat 
 $\Delta C(H) \equiv C(H) -C(0)$, and (c) temperature derivative 
 of the resistivity $d\rho/dT$ and the T-linear coefficient of
 the thermopower $S/T$. The inset shows the magnetic field dependence
 of the specific heat.}
 \label{fig5}
\end{figure}
\subsection{Phase Transition at 22 K}
As shown in the previous section,
the Cu substitution causes the phase transition at 22 K, 
which is probed by the jump of the specific heat,
the dip in the thermopower, and the kink in the resistivity. 
Figure \ref{fig5}(a) shows the specific heat for the $x=0.2$ sample
(the same data in Fig. \ref{fig3}(a)) as a function of temperature 
in linear scale in order to see the 22-K anomaly clearly.
One thing to point out is that the entropy change 
of this transition is surprisingly small.
As shown in Fig. \ref{fig5}(a), we estimated the entropy change by 
the area surrounded with $C/T$ and the dotted line,
which is approximately 77 mJ/Kmol, 
corresponding to 0.01$k_B$ per Co.
Actually only 5\% of Co$_3$O$_4$ impurity 
exhibits a specific heat jump of the same order at 30 K.

There are two possibilities for the origin of the small entropy change.
One is that the 22-K transition is something related to the
impurity phase of the order of 1\%.
Although we cannot exclude this possibility completely,
we will take the other possibility that 
the small entropy change is an intrinsic nature in bulk,
because (i) the field dependence of $C/T$
is different between the 22-K transition and 
the magnetic transition in Co$_3$O$_4$ at 30 K as shown in Fig. \ref{fig5}(b),
(ii) a possible impurity phase is a Cu-based magnetic material,
which is inconsistent with no anomaly in $\chi$ at 22 K,
and (iii) the thermopower and the resistivity systematically
change at the same temperature.
The most familiar phase transition accompanied by a small entropy change
is perhaps superconducting transition.
More generally, off-diagonal long range order 
induces a small entropy change of the order of $Nk_BT/E_F$.

Figure \ref{fig5}(c) shows the $T$-linear term of the thermopower ($S/T$)
and the temperature derivative of the resistivity $d\rho/d T$,
both of which are inversely proportional 
to the Drude weight \cite{comment}.
Their temperature dependences are quite similar to each other,
where the magnitude increases up to almost twice below 22 K.
This indicates that the Drude weight decreases by 50\% at low
temperatures, implying the existence of a (pseudo)gap.
As an off-diagonal long-range order with a gapped state,
one would think of charge density wave (CDW) or spin density wave (SDW).
The calculated Fermi surface \cite{singh}
of the $a_{1g}$ band is hexagon-like, 
which is unstable against CDW or SDW formation with the nesting vector
along the $\Gamma$-K direction.
We think that the 22-K transition is SDW-like,
because CDW is insensitive to magnetic field \cite{coleman}.
Actually we can find many similarities between the 22-K transition and  SDW
transition: 
The entropy change is observed to be quite small 
in Cr \cite{CrReview}, YbBiPt \cite{fisk}
and (TMTSF)$_2$PF$_6$ \cite{coroneus}.
The resistivity shows a hump \cite{CrReview,fisk}, and
the thermopower shows a dip at the transition\cite{choi}.

It is not surprising that the 22-K transition has little effects on 
the magnetic susceptibility.
Since an SDW state is an antiferromagnetically ordered state,
the magnetic excitation is gapless in principle.
In fact, the SDW state of Cr exhibits a very tiny (1-2\%) change
in the susceptibility at the transition temperature \cite{CrReview},
while it causes a clear hump in the resistivity \cite{adachi}.
The metallic conduction below 22 K, implies that a part 
(approximately 50\%) of the Fermi surface remains,
which could smears the SDW transition.  
To clarify the nature of the transition, 
a local magnetic probe such as NMR or $\mu$SR should be employed.

\begin{figure}
 \begin{center}
  \includegraphics[width=8cm,clip]{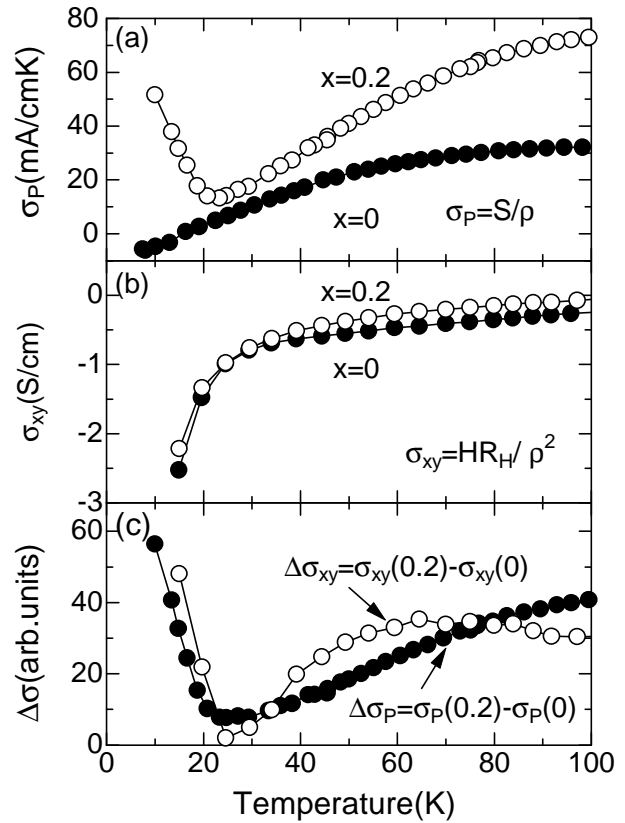}
 \end{center}
 \caption{(a) Peltier conductivity $\sigma_P = S/\rho$ 
 (b) Hall conductivity $\sigma_{xy} = H R_H/\rho^2$ for
 the $x=0$ and $x=0.2$ samples
 (c) the relative change in $\sigma_P$ and $\sigma_{xy}$
 from $x=0$ to 0.2}
 \label{fig6}
\end{figure}
\subsection{Effects on the Hall coefficient and thermopower}
Next we will discuss how we can understand the Cu substitution 
effects on $R_H$ and $S$.
We should note here that the sum rules of transport parameters 
for a multiband system are expressed in the form of conductivities, 
not in the form of $R_H$ or $S$.
Let us denote the conductivities for the $a_{1g}$ 
and $e_g+a_{1g}$ bands as $\sigma^a$ and $\sigma^e$,
respectively.
The total conductivity $\sigma$ is then written as 
\begin{equation}
 \sigma = \sigma^e + \sigma^a 
\end{equation}
Similarly, the total Hall conductivity $\sigma_{xy}$
and the total Peltier conductivity $\sigma_P$
are written as
\begin{eqnarray}
 \sigma_{xy} &=& \sigma_{xy}^e  + \sigma^a_{xy}\\
 \sigma_{P} &=& \sigma_{P}^e  + \sigma^a_{P}
\end{eqnarray}
where the Peltier conductivity \cite{ong} 
is defined as 
$\sigma_P \equiv S \sigma =S/\rho$.

Figure \ref{fig6}(a) shows the temperature dependence of 
$\sigma_P = S/\rho$ for $x=$0 and 0.2.
The Cu substitution enhances the Peltier conductivity 
over the measured temperature range from 4 to 100 K,
indicating that the mobility is enhanced by Cu.
The enhancement below 22 K is more remarkable 
in $\sigma_P$ than in $S$,
which indicates the mobility is rapidly enhanced 
below the 22-K transition.
Figure \ref{fig6}(b) shows the temperature dependence of 
$\sigma_{xy} = HR_H/\rho^2$ for $x=$0 and 0.2.
The complicated change seen in $R_H$ is converted 
into a simpler change in $\sigma_{xy}$.
Although the Cu substitution effects in $\rho$ is quite small,
$1/\rho^2$ term moderates the difference in $R_H$.
One can see that $\sigma_{xy}$ is also increased by Cu
over the temperature range from 15 to 100 K, 
as is similar to the case of $\sigma_{P}$.

Let us assume that the substituted Cu affects only 
the $a_{1g}$ band.
Then a difference between $x=0$ and 0.2
is reduced to a change in $\sigma_P^a$ and $\sigma_{xy}^a$.
Figure \ref{fig6}(c) shows 
$\Delta\sigma_P=\sigma_P(x=0.2)-\sigma_P(x=0)$
$\Delta\sigma_{xy}=\sigma_{xy}(x=0.2)-\sigma_{xy}(x=0)$.
Most unexpectedly, 
the change in the Peltier conductivity $\Delta\sigma_P$
and the change in the Hall conductivity $\Delta\sigma_{xy}$
show nearly the same temperature dependence.
In particular, a clear enhancement below 22 K
indicates that the phase transition causes an equal
impact on $S$ and $R_H$ in the form of 
the Peltier and Hall conductivities.

On the assumption that only the $a_{1g}$ band
is modified by Cu, we will consider
the change in the $a_{1g}$ band in terms of
the carrier concentration $n$, the effective mass $m$,
and the scattering time $\tau$.
Then $\sigma_P^a$ and $\sigma_{xy}^a$ are roughly expressed as
$\sigma_P^a\sim \langle n/m^*\rangle$
and $\sigma_{xy}^a\sim \langle \tau/m^*\rangle$, where
the average of $\langle\cdots\rangle$
is defined as  $(4\pi^3)^{-1}\int (v_F)^2\tau \cdots d^3k$.
A close similarity between $\Delta\sigma_P$ and
$\Delta\sigma_{xy}$ implies that 
$\Delta\langle \tau \rangle$ and $\Delta\langle n \rangle$
are nearly independent of temperature.
The $T$-independent $\Delta\langle \tau \rangle$ means 
the scattering time averaged in the $a_{1g}$ Fermi surface
is dominated by impurity scattering,
which is consistent with the localized picture
of the $a_{1g}$ band.
The positive values of $\Delta\sigma_{xy}$ and
$\Delta\sigma_P$ suggests that the increase of 
$\langle 1/m^* \rangle$.
This indicates that the mass enhancement 
is suppressed (the mobility is enhanced) by Cu
over the measured temperature range,
regardless of the 22-K transition,
which is consistent with the decrease in
$\gamma$ and $\chi$ by Cu.

\begin{figure}
 \begin{center}
  \includegraphics[width=8cm,clip]{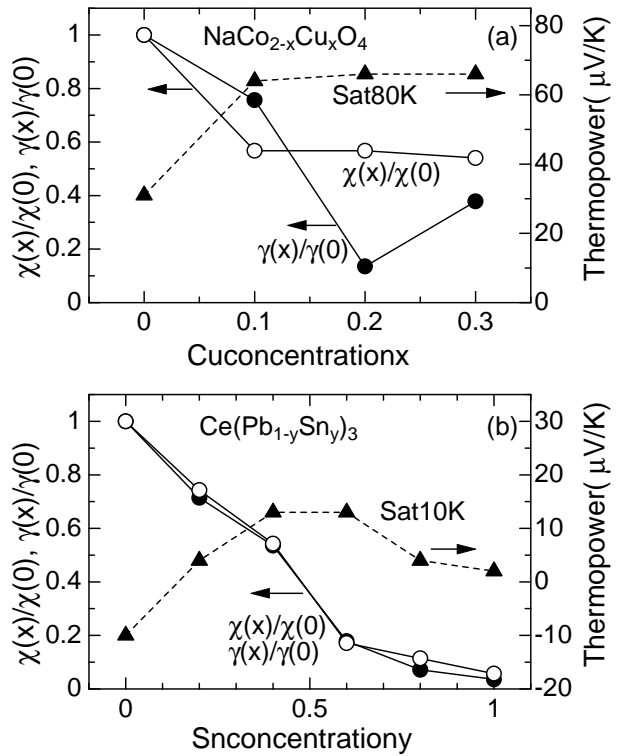}
 \end{center}
 \caption{ (a) Cu dependence of $\chi$, $\gamma$ and S 
 of NaCo$_{2-x}$Cu$_x$O$_4$.
 (b) Sn dependence of $\chi$, $\gamma$ and S of
 Ce(Pb$_{1-y}$Sn$_y$)$_3$. }
 \label{fig7}
\end{figure}
\subsection{Comparison with Ce based compounds}
Based on the heavy-fermion scenario, 
it seems inconsistent that the substituted Cu causes 
the decrease in $\gamma$ (Fig. \ref{fig3}) together 
with the increase in $S$ (Fig. \ref{fig2}). 
As shown in Fig. \ref{fig7}(a), $\gamma$, $\chi$ and $S$ are plotted as
a function of the Cu content $x$.
(Note that $\gamma$ was estimated as $C/T$ at 2 K,
and $\chi$ was estimated as the raw value of $\chi$ at 2 K.)
Although $\gamma$ and $\chi$ decreases with $x$,
$S$ significantly increases with $x$,
where $S \propto \gamma$ is no longer valid.

We should emphasize that the relationship between 
$\gamma$ and $S$ is complicated also in the Ce-based compounds.
Figure \ref{fig7}(b) shows $\gamma$, $\chi$ \cite{lin}
and $S$ \cite{sakurai} for Ce(Pb$_{1-y}$Sn$_y$)$_3$
as a function of the Sn content $y$.
CePb$_3$ is in the heavy fermion regime (low Kondo temperature)
and CeSn$_3$ is in the valence fluctuation regime (high Kondo temperature).
Thus the solid solution between Pb and Sn changes the material 
from the heavy-fermion to the valence-fluctuation compound,
which is evidenced by the fact that $\chi$ and $\gamma$ monotonically
decrease with $y$.
On the other hand, $S$ exhibits complicated $y$ dependence.
$S$ is negative for $y=0$, increases with $y$ up to 0.4,
and eventually decreases from $y$=0.6 to 1.0.

This is intuitively understood as follows.
When the Kondo temperature is sufficiently low as in the case of CePb$_3$, 
the RKKY interaction survives at low temperatures, and
often causes a magnetic transition.
Since the magnetic transition release the entropy of the spin sector,
the entropy per carrier (equivalently the diffusive term
of the thermopower) would be suppressed 
against the fluctuation of the magnetic transition.
On the other hand, when the Kondo temperature 
is high enough, the mass enhancement is severely suppressed
to give a small thermopower again.
Thus the thermopower would take a maximum at an intermediate 
value of the Kondo temperature.
This is indeed what we see in Fig. 7(b),
and similar to the general trend of the 20-K 
thermopower of Ce$M_2X_2$ found 
by Jaccard {\it et al.}\cite{jaccard}

In this context, the NaCo$_2$O$_4$ is located 
near the heavy-fermion regime, 
and the substituted Cu caused the decrease in the mass enhancement
accompanied by the increase in $S$,
which is consistent with the increase in the Peltier and Hall
conductivities seen in the previous subsection.
Although there is no microscopic relationship between 
NaCo$_{2-x}$Cu$_x$O$_4$ and Ce(Pb$_{1-y}$Sn$_y$)$_3$,
a close resemblance in Fig. 7 suggests that the unsubstituted NaCo$_2$O$_4$ 
corresponds to $y\sim0.2$,
while NaCo$_{1.8}$Cu$_{0.2}$O$_4$ 
corresponds to $y\sim 0.4-0.6$.
We further note that the Pd substituted NaCo$_2$O$_4$
shows {\it negative} thermopower below 50 K,
which might correspond to $y< 0.2$ \cite{kitawaki}.

\subsection{Order from disorder}
Although the microscopic theory for the high thermoelectric performance 
of NaCo$_2$O$_4$ is still lacking, the following features are established.
(1) The mixture of Co$^{3+}$ and Co$^{4+}$ in the low spin state
can carry a large entropy of $k_B$log6 \cite{koshibae}.
(2) NaCo$_2$O$_4$ shows no structural, electric, and magnetic
transitions from 2 to 1000 K \cite{ohtaki,fujita}.
(3) From (1)(2), the large entropy cannot be released
through phase transitions, and inevitably adhere to
the conducting carriers to form a ``heavy-fermion''-like electron.

In this respect, NaCo$_2$O$_4$ is very close
to the instability for various phase transitions arising 
from the large entropy per site.
The Cu substitution enhances the instability,
and eventually causes the SDW-like transition at 22 K.
This type of transition is called ``order from disorder''
\cite{tsvelik},
which has been extensively investigated experimentally 
as well as theoretically. 
In other words, instabilities against various phases are competing 
or disordering in NaCo$_2$O$_4$,
and any phase transitions are prohibited down to low temperatures.
This does not mean that NaCo$_2$O$_4$ is far from 
the instability of phase transitions, but rather, 
is very susceptible to various transitions against various perturbations.
In fact, Na$_{1.5}$Co$_2$O$_4$ exhibits a glassy behavior at 3 K 
due to structure instability of the $\gamma$ phase  \cite{takeuchi},
and (Bi,Pb)-Sr-Co-O shows a ferromagnetic transition at 4 K 
due to the lattice misfit \cite{tsukada}.

Among various possible transitions, it is not trivial
whether an SDW-like state is favored by impurities or not.
As an SDW-formation mechanism, we should note here that
SDW and CDW are closely related to the nesting of the 
and the topology of the Fermi surface.
They are a property for a metal,
and occur when the correlation effect is weak enough
to hold one-electron picture based on the band calculation.
As often mentioned in the present paper,
the experimental results consistently 
suggest that Cu suppresses the mass enhancement 
without significant change in the carrier concentration.
If so, the decrease in $\chi$ implies that 
the substituted Cu enhances the screening of 
the magnetic fluctuation, 
which might recover the band picture to cause
the CDW/SDW instability of the $a_{1g}$ Fermi surface.

\section{Summary and Future issues}
In this article, we have discussed the Cu substitution effects on the
thermoelectric and thermodynamic properties of NaCo$_{2-x}$Cu$_x$O$_4$.
The substituted Cu induces a phase transition at 22 K, which is 
characterized by the kink in the resistivity, the hump in the
thermopower, and the jump in the specific heat.
We have analysed the nature of the transition, and finally proposed
a spin-density-wave-like state as a possible origin,
because (i) it accompanies a small entropy change of the order of
10$^{-2} k_B$ per Co, (ii) the transition is sensitive to the 
magnetic field, (iii) the large Fermi surface of the $a_{1g}$
character is gapped.
The impurity induced transition is often called ``order from disorder'',
which implies that phase transitions are somehow suppressed in 
the unsubstituted NaCo$_2$O$_4$.

Above the transition temperature, the thermoelectric properties 
are at least qualitatively compared with those of heavy-fermions
valence-fluctuation compounds,
where mobile holes on the $e_g+a_{1g}$ band and 
the nearly localized holes of the $a_{1g}$ band
correspond to the carrier and the $f$ electrons.
In this analogy,
the substituted Cu increase the interaction between
the $e_g+a_{1g}$ and $a_{1g}$ bands 
to decrease the effective-mass enhancement.

In this article we have reviewed the phenomenology 
of the Cu substitution effects, but failed to address the 
microscopic origin and/or the electronic states of 
the substituted Cu. 
This is because our experiments were concerned only with the thermodynamic and 
transport properties of bulk materials.
Nonetheless we can say that the scattering cross section will be small for 
the $d_{x^2-y^2}$ and $d_{z^2}$ levels of the impurity in NaCo$_2$O$_4$,
because  the valence bands of NaCo$_2$O$_4$ consist of $t_{2g}$.
Thus the substituted Cu (possibly divalent \cite{anno}) 
will not increase $\rho$ seriously,
because Cu$^{2+}$ has the highest occupied orbital of $d_{x^2-y^2}$
that is orthogonal to $t_{2g}$.
In addition, strong Jahn-Teller effects of Cu$^{2+}$ may cause local distortion 
of the CoO$_2$ block, which serves as a kind of chemical pressure 
to increase $S$ \cite{terra2}.
To proceed further, site-selective probes such as NMR,
photoemission, and STM/STS should be employed.

\section*{Acknowledgments}
The authors would like to thank T. Motohashi, H. Yamauchi, N. Murayama,
K. Koumoto, and T. Mizokawa for fruitful discussion.


\begin{thebibliography}{99}
 \bibitem{mahan}
	 G. D. Mahan,
	 Solid State Physics 51, 81 (1998).
 \bibitem{terra}
	 I. Terasaki, Y. Sasago, and K. Uchinokura,
	 Phys. Rev. B56, R12685 (1997)
 \bibitem{fujita}
	 K. Fujita, T. Mochida, and K. Nakamura, 
	 Jpn. J. Appl. Phys. 40, 4644 (2001).
 \bibitem{ohtaki}
	 M. Ohtaki, Y. Nojiri, and E. Maeda,
	 Proc. 19th Int. Conf. Thermoelectrics (ICT2000),
	 ed. D. M. Rowe, p. 190 (Babrow, Wales, 2000).
 \bibitem{li}
	 S. Li, R. Funahashi, I. Matsubara, K. Ueno, and H. Yamada, 
	 J. Mater. Chem. 9, 1659 (1999).
 \bibitem{miyazaki}
	 Y. Miyazaki, K. Kudo, M. Akoshima, Y. Ono, Y. Koike, and
	 T. Kajitani,
	 Jpn. J. Appl. Phys. 39, L531 (2000).
 \bibitem{masset}
	 A. C. Masset, C. Michel, A. Maignan, M. Hervieu, O. Toulemonde,
	 F. Studer, and B. Raveau, 
	 Phys. Rev. B62, 166 (2000).
 \bibitem{funahashi}
	 R. Funahashi, I. Matsubara, H. Ikuta, T. Takeuchi, U. Mizutani, 
	 and S. Sodeoka, 
	 Jpn. J. Appl. Phys. 39, L1127 (2000).
 \bibitem{itoh}
	 T. Itoh, T. Kawata, T. Kitajima, and I. Terasaki, 
	 Proc. 17th Int. Conf. Thermoelectrics (ICT '98), 
	 IEEE, Piscataway, p. 595 (cond-mat/9908039)
 \bibitem{itoh2}
	 T. Itoh and I. Terasaki, 
	 Jpn. J. Appl. Phys. 39, 6658 (2000).
 \bibitem{funahashi2}
	 R. Funahashi and I. Matsubara, 
	 Appl. Phys. Lett. 79, 362 (2001).
 \bibitem{hebert}
	 S. Hebert, S. Lambert, D. Pelloquin, and A. Maignan,
	 Phys. Rev. B74, 172101 (2001).
 \bibitem{terra2}
	 I. Terasaki,
	 Mater. Trans. 42, 951 (2001).
 \bibitem{terra3}
	 I. Terasaki, Y. Ishii, D. Tanaka, K. Takahata, 
	 and Y. Iguchi, 
	 Jpn. J. Appl. Phys. 40, L65 (2001).
 \bibitem{tallon}
	 J. L. Tallon, J. R. Cooper, P. S. I. P. N. de Silva, 
	 G. V. M. Williams, and J. W. Loram,
	 Phys. Rev. Lett. 75, 4114 (1995).  
 \bibitem{fischer}
	 K. H. Fischer,
	 Z. Phys. B 76, 315 (1989) and references therein.
 \bibitem{JH}
	 M. von Jansen and R. Hoppe, 
	 Z. Anorg. Allg. Chem. 408, 104 (1974).
 \bibitem{fouassier}
	 C. Fouassier, G. Matejka, J. Reau, and P. Hagenmuller, 
	 J. Solid State Phys. 6, 532 (1973).
 \bibitem{fukuzumi}
	 Y. Fukuzumi, K. Mizuhashi, K. Takenaka, and S. Uchida,
	 Phys. Rev. Lett. 76, 684 (1996).
 \bibitem{LB1}
	 I. M. Khriplovich, E. V. Kholopov, and I. E. Paukov,
	 J. Chem. Thermodyn. 14, 207 (1982).
 \bibitem{LB2}
	 P. Cossee, 
	 J. Inorg. Nucl. Chem. 8, 483 (1958).
 \bibitem{singh}
	 D. J. Singh, 
	 Phys. Rev. B61, 13397 (2000). 
 \bibitem{mizokawa}
	 T. Mizokawa, L. H. Tjeng, P. G. Steeneken, N. B. Brookes,
	 I. Tsukada, T. Yamamoto, and K. Uchinokura
	 Phys. Rev. B 64, 115104 (2001).
 \bibitem{mizokawa2}
	 T. Mizokawa, L. H. Tjeng, I. Terasaki, H.-J. Lin, C. T. Chen,
	 Synchrotron Radiation Research Center activity report,
	 Taiwan, China (unpublished).
 \bibitem{comment}
	 The diffusive part of the thermopower 
	 in a quasi-two-dimensional conductor is written as 
	 $S/T \propto m/n$, where $n/m$ is the Drude weight 
	 [Mandal et al., Phys. Condens. Matter {\bf 8}, 3047 (1996)].
	 For a polycrystalline NaCo$_2$O$_4$, $\rho$ is
	 roughly written as $aT+b$
	 [Kawata et al., Phys. Rev. B 60, 10584 (1999)],
	 and thus $d\rho/dT \propto m/n$.
 \bibitem{coleman}
	 R. V. Coleman, M. P. Everson, Hao An Lu, A. Johnson 
	 and L. M. Falicov,
	 Phys. Rev. B41, 460 (1990).
 \bibitem{CrReview}
	 E. Fawcett, H. L. Alberts, V. Yu. Galkin, D. R. Noakes, and
	 J. V. Yakhmi,
	 Rev. Mod. Phys. 66, 25 (1996).
 \bibitem{fisk}
	 R. Movshovich, A. Lacerda, P. C. Canfield, J. D. Thompson, and
	 Z. Fisk,
	 Phys. Rev. Lett. 73, 492 (1994).
 \bibitem{coroneus}
	 J. Coroneus, B. Alavi and S. E. Brown,
	 Phys. Rev. Lett. 70, 2332 (1993).
 \bibitem{choi}
	 M. Y. Choi, M. J. Burns, P. M. Chaikin, E. M. Engler 
	 and R. L. Greene,
	 Phys. Rev. B31, 3576 (1985).
 \bibitem{adachi}
	 S. Maki and K. Adachi,
	 J. Phys. Soc. Jpn. 46, 1131 (1979).
 \bibitem{ong}
	 Z. A. Xu, N. P. Ong, Y. Wang, T. Kakeshita and S. Uchida,
	 Physica C 341-348, 1713 (2000).
 \bibitem{lin}
	 C. L. Lin, J. E. Crow, P. Schlottmann, and T. Mihalisin,
	 J. Appl. Phys. 61, 4376 (1987).
 \bibitem{sakurai}
	 J. Sakurai, H. Kamimura, and Y. Komura,
	 J. Mag. Mag. Mater. 76\&77, 287 (1988).
 \bibitem{jaccard}
	 D. Jaccard, K. Behnia, and J. Sierro
	 Phys. Lett. A163, 475 (1992).
 \bibitem{kitawaki}
	 R. Kitawaki and I. Terasaki,
	 (in preparation)
 \bibitem{koshibae}
	 W. Koshibae, K. Tsutsui, and S. Maekawa,
	 Phys. Rev. B 62, 6869 (2000)
 \bibitem{tsvelik}
	 A. M. Tsuvelik,
	 ``Quantum field theory in condensed matter physics'',
	 (Cambridge University Press, 1995, Cambridge) p.174.
 \bibitem{takeuchi}
	 T. Takeuchi, M. Matoba, T. Aharen, and M. Itoh,
	 Physica B (in press)
 \bibitem{tsukada}
	 I. Tsukada, T. Yamamoto, M. Takagi, T. Tsubone, S. Konno, 
	 and K. Uchinokura, 
	 J. Phys. Soc. Jpn. {\bf 70},834 (2001).
 \bibitem{anno}
	 H. Anno, 
	 Proc. 20th Int. Conf. Thermoelectrics (ICT 2001)
\end{thebibliography}
\end{document}